\documentclass[fleqn,2pt]{article}
\usepackage{anysize}
\marginsize{2.5cm}{2cm}{2cm}{2cm}

\usepackage{amsmath,amsfonts,amssymb,epsfig}
\numberwithin{equation}{section}

\usepackage{cancel}
\usepackage{cite}

\def\tr{\mathrm{tr}}

\def\beq{\begin{equation}}
\def\eeq{\end{equation}}
\def\bal{\begin{align}}
\def\eal{\end{align}}

\def\2b2[#1,#2][#3,#4]{\left( \begin{array}{cc} #1 & #2 \\ #3 & #4 \end{array}
\right)}
\def\3b3[#1,#2,#3][#4,#5,#6][#7,#8,#9]{\left( \begin{array}{ccc} #1 & #2 #3 \\
#4 & #5 & #6\\#7&#8&#9\end{array} \right)}

\newcommand{\C}[1]{\mathcal{#1}}

\usepackage[]{graphicx}

\author{Karim Benakli\footnote{kbenakli@lpthe.jussieu.fr}}
\date{}
\title{\vspace{-3cm}
Dirac Gauginos: A User Manual}
\begin{document}
\maketitle
\vspace{-1cm}
\begin{center}
\emph{Laboratoire de Physique Th\'eorique et Hautes Energies,  CNRS, UPMC
Univ Paris VI
Boite 126, 4 Place Jussieu, 75252 Paris cedex 05, France}
\end{center}

\begin{abstract}
The issue of a Majorana, Dirac or  pseudo-Dirac mass for gauginos 
must not be reduced to a question of an extension of the Minimal Supersymmetric Standard Model
by extra states, parameters and phenomenological implications.  On the contrary,
it is intimately related to the fundamental issue of the realization of new symmetries in nature,  
 $R$-symmetries. We present here a very dense compilation of the main features  of models with (pseudo-)Dirac
gauginos.
\end{abstract}
\maketitle                   





\section{Introduction: identification of $R$-symmetry breaking terms}
Symmetries rule the fundamental law of nature. They provide the guideline for building
theoretical models designed to describe the world. While two degrees of freedom are sufficient to describe a
neutral massive fermion, four degrees of freedom are necessary if the fermion carries charges under a conserved
symmetry. More precisely, two elementary 
Majorana fermions have then to pair up to produce a Dirac fermion with the same
mass.

If supersymmetry is discovered at collider experiments, there is no doubt that
the question of the nature of gauginos masses will then become of central importance.
This is because 
going from Majorana to Dirac or  pseudo-Dirac masses is not just a question of 
extra states and parameters but also a signal the presence of $R$-symmetry, at
least in a sector of the theory. $R$-symmetries  are special as they do not commute with 
the supersymmetry supercharges, and thus acts differently on the components of  superfields.  

In the framework of global supersymmetric models considered here, $R$-symmetry
appears as a continuous
symmetry (which can be broken to a discrete subgroup).  It can not be
spontaneously broken at the electroweak scale with a generic vacuum expectation
value (vev), as this would lead to a massless $R$-axion with a coupling
insufficiently suppressed to have evaded early
discovery. There remain two options: either it is conserved or explicitly
broken. 

In order to quantify the required size of $R$-symmetry breaking, one
needs to identify the minimal set of operators that violate the symmetry.
\begin{enumerate}
 \item First, it is usual to consider that $R$-symmetry is  broken in the
Minimal Supersymmetric extension of the
Standard Model (MSSM) by Majorana gaugino masses. We can however use instead
Dirac masses for the gauginos, pairing them with additional states in adjoint
representations: a singlet $\mathbf{S}$ for $U(1)_Y$, a
triplet $\mathbf{T}$ for $SU(2)_w$ and an octet $\mathbf{O_g}$ for $SU(3)_c$, and preserve 
$R$-symmetry. 

 \item Second, in $N=1$ supergravity, the gravitino mass required in flat
space-time breaks $R$-symmetry. 
Again, this can be avoided by giving a Dirac mass for the
gravitino. This requires the gravitational multiplet to be in extended $N=2$
representations. 
To illustrate such a scenario, consider that the $N=1$ gauge and matter fields
appear on 3-branes. These are
localized in a bulk having one flat extra dimension of radius $R$.  Then a
Dirac gravitino mass of size $1/2R$, and preserving $R$-symmetry,  is obtained
when the $N=2$ supergravity is  broken through a Scherk-Schwarz mechanism. 

 \item Finally, the  presence of the  $\mu$ and $B\mu$ mass terms for
the Higgs doublets is incompatible with an $R$-symmetry. This makes it  difficult  to
arrange a satisfactory  
electroweak symmetry breaking  which preserves $R$-symmetry. We shall therefore
consider the Higgs sector
 as the main source of $R$ -symmetry breaking.

\end{enumerate}

\section{Dirac versus Majorana gauginos}

Let us point out some of the main features and challenges raised by Dirac
gauginos.
\begin{itemize}

\item \textbf{Minimal  content:}  
The adjoint representation $\mathbf{S}$,
$\mathbf{T}$ and  $\mathbf{O_g}$
 represent the minimal set of superfields to be added to the MSSM to allow 
(pseudo-)Dirac masses. Such
states can be remanents of 
  an $N=2$ extended supersymmetric structure at higher energies, that is broken by
interactions with the $N=1$ matter sector 
\cite{Fayet:1978qc}. 
Therefore, an experimental discovery of such new states might 
be viewed as an indication for the extended supersymmetry origin of the model.

 \item \textbf{Softness:} As Majorana masses, Dirac ones can
be soft \cite{Fayet:1978qc}. 
However, to achieve this important property, one needs the presence of 
new trilinear  terms \cite{Fayet:1978qc,Polchinski:1982an}. In fact, 
  Dirac gaugino masses can be described with superfields by the Lagrangian:
\begin{align}
\mathcal{L}^{Dirac}_{gaugino}=& \int  d^2\theta  \left[   \sqrt{2}
\textbf{m}^\alpha_{1D} \mathbf{W}_{1\alpha} \mathbf{S}  
+  2\sqrt{2} \textbf{m}^\alpha_{2D} \textrm{tr}(\mathbf{W}_{2 \alpha}
\mathbf{T}) +  2\sqrt{2} \textbf{m}^\alpha_{3D} \textrm{tr}(\mathbf{W}_{3
\alpha} \mathbf{O_g})   \right]  +h.c. 
\label{Newdiracgauge}
\end{align}
where we have introduced  spurion superfields $\textbf{m}_{\alpha iD} =
\theta_\alpha m_{iD}$. The  integration over 
the Grasmannian coordinates leads then to
\beq
\int d^2 \theta 2\sqrt{2} m_D \theta^\alpha \tr (W_\alpha \Sigma) \supset - m_D
(\lambda_a \psi_a) + \sqrt{2} m_D \Sigma_a D_a
\eeq
Then with $D^a_b = - g_b \phi^\dagger_i R^a_b (i) \phi_i $ (where
$R_b^a (i) $ is the $a^{th}$ generator of the group $b$ in the representation of
field $i$, and $R_Y^b (i) = Y (i)$ for the hypercharge) we find $\C{L} \supset -
m_{bD} \sqrt{2} 
 g_b \Sigma_a \phi^\dagger R^a_b \phi$.
The new terms are proportional to the Dirac masses. 

Simple scaling 
arguments indicate that such soft terms are in fact even softer than Majorana 
ones; they lead to finite contributions i.e. without logarithmic divergences
\cite{Fox:2002bu}. Note however, 
 that in realistic models,  as the  Higgs sector
induces a small $R$-symmetry breaking, higher orders radiative corrections might
lead to resurgence of 
logarithmic divergences.

\item \textbf{Unification:} 
New states can  be introduced at some mass scale 
to modify at will the running of gauge couplings, if one allows
their number and  charges to remain arbitrary. Though it can
no more be viewed as a prediction, unification of couplings can  be engineered.
The non-singlet extra states have dramatic effects on  the running of the
standard model gauge couplings. Already, the adjoint octet  spoils the
asymptotic freedom of the strong interaction, 
making the one-loop beta function coefficient vanish and raises a difficulty. In order to keep 
the quantitative predictability of the theory, it is essential to ensure that
the couplings remain perturbative in all 
the energy range where the theory is studied. This puts strong constraints on
additional states 
that are needed to achieve unification. It makes not easy  to simultaneously provide non-gravitational 
messengers of supersymmetry breaking. 
Still, such a possibility can  be realized as shown in \cite{Benakli:2010gi}.

\item \textbf{Generating gaugino masses:} In a top-down approach, one hopes
that  the gaugino masses can originate from 
dynamical breaking that leads to either $D$-terms $<\!\!D\!\!>\neq 0$, $R$-preserving
$F$-terms  $<\!\!F\!\!> \neq 0$ or both. This would
allow to shed some light on the origin of hierarchies in the sparticles
spectrum.

In a gauge mediation type scenario, new states are introduced at  a mass scale
$M_{mess}$  to serve as mediators of the breaking, 
and allow to generate Dirac gaugino masses. The gaugino masses are of order \cite{Benakli:2008pg}:
\begin{align}
m^{(<D>)}_{D}  \sim \frac{<D>}{M_{mess}}  \qquad {\rm and} \qquad 
m^{(<F>)}_{D}\sim \frac{<F>^2}{M^3_{mess}} 
\end{align}
with the corresponding one-loop factors. Note that to obtain sizable contributions
to all soft terms from the second case,  
the scale of the $F$-term needs to be close to the messengers scale, therefore
both low. This leads to a loss
of perturbative nature of the gauge couplings at intermediate
scales\cite{Amigo:2008rc,Abel:2011dc}. It is therefore preferable to rely on $D$-term
contributions. In which case, the Dirac mass can be more easily 
extracted from computing the appropriate $U(1)$ mixing kinetic functions
\cite{Benakli:2009mk}.
Note that the same messenger states can be used to restore unification of the gauge
couplings \cite{Benakli:2010gi}. 

In gravity mediation scheme, one needs to consider models where the spontaneous
supersymmetry breaking results in  a Dirac gravitino of mass $m_{3/2}$. The
mediation of the breaking to the observable sector gauginos can not generate
Majorana gaugino masses as they are protected by $R$-symmetry, but can 
induce Dirac masses. From dimensional analysis, one can easily see that the
expected size for gaugino and scalar masses is generically to be of order $m_{3/2}$.

As a special case, let us mention the possibility of a set-up, where the Dirac
gravitino is the messenger that communicates supersymmetry breaking to the
observable sector through radiative effects. 
The Dirac gaugino  mass is now generated at two-loops and found to be of order $\frac{m_{3/2}^2}{M_{Planck}}$.
The power suppression (square) arises from the fact that only one of the two degenerate Majorana 
gravitinos interact with the observable sector.
The scalar soft terms are also  of the same order in $m_{3/2}$. 

\item \textbf{Adjoint scalars soft masses:} The generation of a soft mass for
the adjoint scalars turns out to be less trivial than one expects 
in the case of gauge mediation. This is because the simplest interactions
between the DG-adjoints and the messengers, as the Yukawa couplings descending
from $N=2$ supersymmetric gauge structure, lead to tachyonic masses.
Historically, this was the main reason for abandoning the Dirac gaugino scenario in 
 \cite{Fayet:1978qc}. To avoid such a result, the required forms of the
adjoint-messengers interactions have  been fully classified in 
\cite{Benakli:2008pg}.

Two other other important constraints need to be considered when dealing with these masses.
One is that the $U(1)$ adjoint scalar is a singlet, and therefore
subject to  more stringent constraints, as to avoid unwanted tadpoles
\cite{Benakli:2010gi}. The second, is that the $SU(2)_w$ adjoint needs to be sufficiently massive 
to avoid getting a large vev after electroweak symmetry breaking.

\item \textbf{Scalars soft masses:}  At leading order, at  which $R$-symmetry breaking effects are neglected, 
the scalar masses appear to be finite \cite{Fox:2002bu}. They are induced  by radiative corrections from
Dirac gaugino masses. This feature has been used in \cite{Kribs:2007ac} to partially 
ease the supersymmetric flavor problem, as it allows to keep radiatively stable a desired 
 hierarchy between sparticles.

Note that, contrary to the gaugino case, it is now the  $D$-term
contribution to the scalar masses that appears to be dangerously small, as it
might lead to a too light selectron, for example. Therefore, all together, a preferable
pattern for model building is to combine the $D$-term induce gaugino mass with
$F$-term induced scalar masses \cite{Benakli:2010gi,Davies:2011mp}.

\item \textbf{RGE running of masses:} The renormalization group equations
(RGE) for running masses are different from the MSSM ones. Contrary to the Majorana case,
the Dirac gaugino masses do not enter at one-loop in the RGE of squarks and
sleptons \cite{Benakli:2009mk}, leading to new patterns in the generated
spectra.

\item \textbf{Electroweak sector:}  
Our simple model for the Higgs sector has a superpotential
\begin{eqnarray}
W^{(s)} &=& \mu \mathbf{H_u\!\cdot\! H_d }   + \lambda_S \mathbf{SH_u\!\cdot\!
H_d}  + 2  \lambda_T \mathbf{H_d\!\cdot\! T H_u} + \frac {M_S}{2}\mathbf{S}^2 +
\frac{\kappa}{3}
\mathbf{S}^3 + M_T \textrm{tr}(\mathbf{TT}) 
\label{NewSuperPotential}
\end{eqnarray}
with supersymmetry breaking described by the soft masses
\begin{eqnarray}
\label{potential4}
- \Delta\mathcal{L}^{soft} &= & m_{H_u}^2 |H_u|^2 +
m_{H_d}^2 |H_d|^2
 + [\tilde{B}{\mu} H_u\cdot H_d + h.c. ] \nonumber \\
 &+&  m_S^2  |S|^2 + \frac{1}{2} B_S
(S^2 + h.c.)  + 2 m_T^2 \textrm{tr}(T^\dagger T) + B_T (\textrm{tr}(T T)+ h.c.)
\nonumber \\  &&+ 
A_S  \lambda_S SH_u\cdot H_d +  2 A_T \lambda_T H_d \cdot T H_u +
\frac{1}{3} \kappa  A_{\kappa} S^3
\label{Lsoft-DGAdjoint}
\end{eqnarray}
where as in \cite{Nelson:2002ca,Antoniadis:2006uj} we have chosen to allow $R$-symmetry to be broken by
a $B\mu$ mass. In addition, we have included another source of $R$-symmetry through $\kappa \neq 0$ term.

The singlet cubic interaction can be used to generate an effective $\mu$-term
$\tilde{\mu}$ through the singlet vev.
As $\kappa$ breaks the $R$-symmetry, it induces both a $B\mu$ term of order 
$\tilde{B}{\mu} \simeq\frac{\kappa}{\lambda_S}  \tilde{\mu}^2$ as well as a
singlino Majorana mass of order  
$M'_1 \simeq\frac{{2} \kappa}{\lambda_S}  \tilde{\mu}$.

An important feature of the new electroweak sector is  the presence of a term $\lambda_S
\mathbf{SH_u\!\cdot\! H_d}$ as stressed in \cite{Antoniadis:2006uj}. This allows one to
increase the tree-level mass of the lightest Higgs and be above the LEP bound
without appealing to the large stop-top loop corrections \cite{Benakli:2011kz}.

We remind that to satisfy the constraints from by the electroweak
precision tests,  the triplet vev must remain small.
This requires the scalar triplet mass to be an order of magnitude larger than
the electroweak scale, 
which can be achieved easily in models of gauge mediation
\cite{Benakli:2010gi}. 

\item \textbf{Collider signatures:} The main collider signature is the
production of scalar octets at LHC. It requires 
the octet mass to be of the order of 1 TeV 
and needs use of peculiar cuts in analysis of LHC data, as described in \cite{Choi:2008ub}.

\item \textbf{Dark matter:}  On one hand, the LSP candidate can be a neutralino. 
Unless there is a remanent $N=2$ structure at low energies, this is a 
Majorana particle\cite{Belanger:2009wf} nearly degenerate with the state made by
the NLSP, which opens up new co-annihilation 
channels that help to get the correct dark matter relic abundance. On the other hand, the LSP
can be  the gravitino. When this appears as 
a Dirac or pseudo-Dirac state,  the usual computation of the relic
abundance is modified by  the interactions of the ``hidden Majorana gravitino''
 both with the observable one and with the (hidden) matter fields  \cite{Belanger:2009wf,Benakli}.
\end{itemize}

\section*{Acknowledgement}
I am grateful to the organizers of the 10th Hellenic School on Elementary
Particle Physics and Gravity, Corfu 2010  for their kind hospitality and to
I. Antoniadis, A. Delgado, M. D. Goodsell and M. Quir\`os for illuminating
discussions during collaborations on different parts of the material presented
in these proceedings. This work is supported in part by the European contract
``UNILHC'' PITN-GA-2009-237920.

\end{document}